\documentclass[12pt]{article}
\usepackage{epsfig}
\usepackage{amssymb}
\usepackage{amsmath}
\usepackage{amsfonts}
\usepackage{graphicx}
\usepackage{mathrsfs}
\usepackage[dvips]{color}
\usepackage{multirow}

\newcommand{\bsigma}{\boldsymbol{\sigma}}

\newcommand{\R}{\mathbb{R}}
\newcommand{\C}{\mathbb{C}}

\newcommand{\fz}{\mathfrak{z}}

\newcommand{\fK}{\mathfrak{K}}

\newcommand{\bbe}{\mathbf{e}}

\newcommand{\bk}{\mathbf{k}}

\newcommand{\bbr}{\mathbf{r}}

\newcommand{\bzero}{\mathbf{0}}

\newcommand{\bH}{\mathbf{H}}
\newcommand{\bI}{\mathbf{I}}

\newcommand{\bM}{\mathbf{M}}

\newcommand{\bU}{\mathbf{U}}

\newcommand{\cF}{\mathcal{F}}

\newcommand{\cK}{\mathcal{K}}

\newcommand{\be}{\begin{equation}}
\newcommand{\ee}{\end{equation}}
\newcommand{\bea}{\begin{eqnarray}}
\newcommand{\eea}{\end{eqnarray}}

\newcommand{\ed}{\end{document}}

\newcommand{\bi}{\begin{itemize}}
\newcommand{\ei}{\end{itemize}}

\newcommand{\bce}{\begin{center}}
\newcommand{\ece}{\end{center}}

\newcommand{\sD}{\mathscr{D}}

\newcommand{\sF}{\mathscr{F}}

\newcommand{\sR}{\mathscr{R}}

\newcommand{\sT}{\mathscr{T}}

\newcommand{\bcK}{{\boldsymbol{\cK}}}

\begin{document}


\title{Potentials with Identical Scattering Properties\\
Below a Critical Energy}

\author{Farhang Loran\thanks{E-mail address: loran@cc.iut.ac.ir}
~and Ali~Mostafazadeh\thanks{E-mail address:
amostafazadeh@ku.edu.tr}\\[6pt]
$^*$Department of Physics, Isfahan University of Technology, \\ Isfahan 84156-83111, Iran\\[6pt]
$^\dagger$Departments of Mathematics and Physics, Ko\c{c} University,\\  34450 Sar{\i}yer,
Istanbul, Turkey}

\date{ }
\maketitle

\begin{abstract}
A pair of scattering potentials are called $\alpha$-equivalent if
they  have identical scattering properties for incident plane waves
with wavenumber $k\leq\alpha$ (energy $k^2\leq\alpha^2$.) We use a
recently developed multidimensional transfer-matrix formulation of
scattering theory to obtain a simple criterion for
$\alpha$-equivalence of complex potentials in two and three
dimensions.



\end{abstract}

One of the basic results of potential scattering is the uniqueness of the solution to the inverse scattering problem. This means that under fairly general conditions on the scattering potential, the information about its scattering properties for incident waves of all wavenumbers determines the potential in a unique manner \cite{IS}. This is a mathematical result with a rather limited practical impact, because for a realistic scattering problem the scattering data can be collected only for a finite range $\sR$ of values of the incident wavenumber. The application of the inverse scattering prescriptions to such an incomplete scattering data cannot yield a unique potential, because the information about the scattering properties outside $\sR$ is missing. What one can hope for is to identify the class of potentials whose scattering properties coincide in $\sR$. To the best of our knowledge a complete solution of this partial inverse scattering problem is still out of reach. The purpose of the present article is to offer a rather general solution for this problem in two and three dimensions for the cases that $\sR$ is a finite interval of the form $(0,\alpha]$ and the potentials are allowed to take complex values.

Among the best known tools for carrying out scattering calculations in one dimension is the notion of transfer matrix \cite{abeles,thompson,levesque,hosten,pereyra,griffiths,yeh,prl-2009,sanchez,bookchapter}. This is a complex $2\times 2$ matrix whose entries determine the reflection and transmission amplitudes of the potential \cite{prl-2009}. A remarkable property of the transfer matrix which makes it into an effective tool for performing scattering calculations is its composition property \cite{bookchapter}; if we slice a scattering potential $v(x)$ into a finite number of pieces, $v_1(x),v_2(x),\cdots,v_n(x)$, such that $v(x)=\sum_{\ell=1}^n v_\ell(x)$ and the support of $v_\ell(x)$ lies to the left of that of $v_{\ell+1}(x)$, then the transfer matrix of $v(x)$ takes the form $\bM_n\bM_{n-1}\cdots\bM_1$, where $\bM_\ell$ is the transfer matrix of $v_\ell(x)$. Ref.~\cite{pra-2014a} offers a curious explanation for this behavior by identifying the transfer matrix of $v(x)$ with the the S-matrix of a fictitious nonunitary two-level quantum system. This in turn paves the way for the development of a comprehensive transfer-matrix formulation of scattering theory in two and three dimensions \cite{pra-2016}. In exploring the applications of this formulation in two-dimensions we were led to the following surprising observations:
    \begin{enumerate}
    \item Given a wavenumber scale $\alpha$, there is an infinite class of scattering potentials in two-dimensions that are invisible for every incident plane wave with wavenumber $k\leq\alpha$, \cite{ol-2017}. In other words, for these wavenumbers, they have the same scattering properties as the zero potential.
    \item Consider potentials of the form
        \begin{align}
        &v(x,y)=\delta(x)\sum_{n=-\infty}^\infty \fz_n\,e^{i n\alpha_1 y},
        &&v_N(x,y)=\delta(x)\sum_{n=-N}^N \fz_n\,e^{i n\alpha_1 y},
        \label{eq1}
        \end{align}
where $\delta(x)$ denotes the Dirac $\delta$ function, $\fz_n$ are a sequence of real or complex numbers for which the Fourier series $\sum_{n=-\infty}^\infty \fz_n\,e^{i n \alpha_1 y}$ converges, $\alpha_1$ is a positive real parameter, and $N$ is a nonnegative integer. Then $v$ and $v_N$ have identical scattering properties for incident wavenumbers $k<\alpha_N$, where $\alpha_N:=\alpha_1(N+1)/2$, \cite{jpa-2018}.
    \end{enumerate}
Both of these concern different potentials sharing the same scattering features in an extended range of wavenumbers.

Let us use the term ``$\alpha$-equivalence'' to refer to the property of having identical scattering properties for all incident waves having a wavenumber $k\leq\alpha$. The purpose of the present article is to give a simple criterion for the $\alpha$-equivalence of scattering potentials in two and three dimensions. Our main tool is the multidimensional transfer-matrix formulation of scattering theory that we have developed in Refs.~\cite{pra-2016,prsa-2016}. We therefore begin with a brief review of this formulation.

First, we consider potential scattering in two dimensions.

Let $v(x,y)$ be a possibly complex-valued scattering potential. We use the symbol $\tilde v(x,\fK_y)$ to denote the Fourier transform of $v(x,y)$ with respect to $y$, i.e.,
    \[ \tilde v(x,\fK_y):=\cF_{\fK_y}\{v(x,y)\}:=\int_{-\infty}^\infty dy\,e^{-i\fK_y y}v(x,y),\]
introduce the function spaces:
    \[\sF_k^d:=\Big\{\,\phi:\R\to\C^d\,\Big|\, \phi(p)=0~{\rm for}~p\notin(-k,k)\,\Big\},\]
with $d=1,2$, and use $\tilde v(x,\fK_y)$ to define an integral operator, $v(x,i\partial_p):\sF_k^1\to\sF_k^1$, via
    \be
    [v(x,i\partial_p)\phi](p):=\left\{\begin{array}{cc}
    \displaystyle \frac{1}{2\pi}\int_{-k}^k dq\: \tilde v(x,p-q)\phi(q) &{\rm for}~|p|<k,\\
    0 &{\rm for}~|p|\geq k.\end{array}\right.
    \label{v-op}
    \ee

Now, consider the quantum system whose state vectors $\Phi$ belong to $\sF_k^2$ and whose dynamics is determined by the Schr\"ondinger equation,
$i\partial_x\Psi(x)=\bH(x)\Psi(x)$, where $x$ plays the role of `time,' $\bH(x):\sF_k^2\to\sF_k^2$, as defined by
    \be
        [\bH(x)\Phi](p):=\frac{1}{2\varpi(p)}\: e^{-i\varpi(p)x{\bsigma}_3}
        v(x,i\partial_p)\,{\bcK}\,e^{i\varpi(p)x{\bsigma}_3}\Phi(p),
        \label{H=}
        \ee
is the Hamiltonian operator,
    \begin{align}
    &\varpi(p):=\sqrt{k^2-p^2},
    &&{\bcK}:={\bsigma}_3+i{\bsigma}_2=\left[\begin{array}{cc}
    1 & 1\\
    -1 & -1\end{array}\right],
    \label{omega=}
    \end{align}
and ${\bsigma}_i$ are the Pauli matrices \cite{pra-2016}. Let $\bU(x,x_0)$ denote the evolution operator for this system. By definition, it satisfies:
    \begin{align}
    &i\partial_x \bU(x,x_0)=\bH(x)\bU(x,x_0),
    &&\bU(x_0,x_0)=\bI,
    \label{sch-eq-U}
    \end{align}
where $x_0$ is an initial value of $x$, and $\bI$ is the $2\times 2$ identity matrix. It is customary to express the solution of (\ref{sch-eq-U}) as the time-ordered exponential \cite{weinberg}:
    \[\bU(x,x_0)=\sT\exp\left[-i\int_{x_0}^x dx'\,\bH(x')\right],\]
where $\sT$ stands for the time-ordering operation with $x$ playing the role of time.

The transfer matrix for the potential $v(x,y)$ is given by
    \be
    \bM=\bU(\infty,-\infty)=\exp\left[-i\int_{-\infty}^\infty dx\,\bH(x)\right].
    \label{M=}
    \ee
It is a $2\times 2$ matrix $\bM$ with operator entries $M_{ij}$ acting in $\sF_k^1$, \cite{pra-2016}. This notion of transfer matrix shares the basic properties of its well-known one-dimensional analog \cite{prl-2009,sanchez,bookchapter}. In particular, it obeys a similar composition rule and encodes all the information about the scattering features of the potential \cite{pra-2016}.

To elucidate the relevance of the transfer matrix (\ref{M=}) to the scattering problem for the potential $v(x,y)$, we consider scattering solutions, $\psi^l(x,y)$ and $\psi^r(x,y)$, of the Schr\"odinger equation, $\left[-\nabla^2+v(x,y)\right]\psi(x,y)=k^2\psi(x,y)$, that are respectively associated with an incident wave whose source resides at $x=-\infty$ and $x=+\infty$. We refer to these as ``left-incident'' and ``right-incident'' waves and denote the corresponding scattering amplitude by $f^l(\theta)$ and $f^r(\theta)$, respectively. This means that $\psi^{l/r}(x,y)$ satisfies the following asymptotic boundary condition.
    \bea
        \psi^{l/r}(\bbr)&=&e^{i\bk^{l/r}_0\cdot
    \:\bbr}+\sqrt{\frac{i}{kr}}\,e^{ikr} f^{l/r}(\theta)
    ~~~~{\rm as}~r\to\infty,
        \label{psi-left}
     \eea
where we have identified $(x,y)$ with the position vector: $\bbr=x\,\hat\bbe_x+y\,\hat\bbe_y$, $\hat\bbe_j$ is the unit vector along the $j$-axis, $\bk_0^{l/r}$ is the wave vector for the left/right incident wave (with $|\bk_i^{l/r}|=k$ and the $x$-component of $\bk_0^{l/r}$ taking positive/negative values), and $(r,\theta)$ are polar coordinates of $\bbr$. The scattering amplitudes $f^{l/r}(\theta)$ turn out to admit the following expression \cite{pra-2016,ol-2017,prsa-2016}.
    \be
    f^{l/r}(\theta)=-\frac{ik|\cos\theta|}{\sqrt{2\pi}}\times \left\{
        \begin{array}{cc}
        T^{l/r}_-(k\sin\theta) &{\rm for}~ \cos\theta<0,\\
        T^{l/r}_+(k\sin\theta) &{\rm for}~  \cos\theta>0,
        \end{array}\right.
        \label{f=}
        \ee
where
    \begin{align}
        &T^l_-(p)=-2\pi M_{22}^{-1}M_{21}\delta(p-p_0),
        \label{Tm-L}\\
        &T^l_+(p)=2\pi\left[M_{11}-1-M_{12}M_{22}^{-1}M_{21}\right]
    \delta(p-p_0),
        \label{Tp-L}\\
    &T^{\rm r}_-(p)=-2\pi \left[1-M_{22}^{-1}\right]\delta(p-p_0),
        \label{Tm-R}\\
        &T^{\rm r}_+(p)=2\pi M_{12}M_{22}^{-1}\delta(p-p_0).
        \label{Tp-R}
        \end{align}
Equations (\ref{f=}) -- (\ref{Tp-R}) reduce the solution of the scattering problem for the potential $v(x,y)$ to the determination of the transfer matrix $\bM$ and the inversion of $M_{22}$, which is an integral operator acting in $\sF^1_k$. Details of the application of this approach for solving specific scattering problems are given in Refs.~\cite{pra-2016,jpa-2018,pra-2017}.

The transfer-matrix formulation of potential scattering in two dimensions provides a convenient framework for the study of invisible potentials. To see this, first we note that according to (\ref{H=}) and (\ref{omega=}) the Hamiltonian operator $\bH(x)$ depends on the wavenumber $k$. In light of (\ref{M=}), this implies that the same holds for the transfer matrix $\bM$. If $\bH(x)$ vanishes for a range of values of $k$, say $\sR$, $\bM$ coincides with the identity operator acting in $\sF^2_k$ for $k\in\sR$. This means that
    \begin{align}
    &M_{11}=M_{22}=I,
    &&M_{12}=M_{21}=O,
    \label{M=inv}
    \end{align}
where $I$ and $O$ respectively label the identity and zero operators acting in $\sF_k^1$. Substituting (\ref{M=inv}) in (\ref{Tm-L}) -- (\ref{Tp-R}) and using the result in (\ref{f=}), we find $T^{l/r}_\pm(p)=0$ and $f^{l/r}(\theta)=0$. Therefore $v(x,y)$ is invisible for $k\in\sR$. In Ref.~\cite{ol-2017}, we employ this argument to prove the following theoem.
    \begin{itemize}
    \item[]{\em Theorem~1:} Let $\alpha$ be a wavenumber scale. Then a scattering potential $v(x,y)$ is invisible for incident waves with wavenumber $k\leq\alpha$, if
    \be
    \tilde v(x,\fK_y)=0~~~{\rm for}~~~\fK_y\leq2\alpha.
    \label{condi-inv}
    \ee
    \end{itemize}
The proof of this theorem relies on the observation that whenever (\ref{condi-inv}) holds, the right-hand side of (\ref{v-op}) vanishes for $k\leq\alpha$. Therefore, $\bH(x)$ vanishes, $\bM$ coincides with the identity operator, and $v(x,y)$ is invisible for this range of values of $k$.

Now, consider a pair of scattering potentials $v_1(x,y)$ and $v_2(x,y)$, with scattering amplitudes $f^{l/r}_1(\theta)$ and $f^{l/r}_2(\theta)$, transfer matrices $\bM_1$ and $\bM_2$, and the associated Hamiltonians, $\bH_1(x)$ and $\bH_2(x)$. Suppose that for $k\in\sR$, $\bH_1(x)=\bH_2(x)$. Then the transfer matrices $\bM_1$ and $\bM_2$ coincide, and we can use (\ref{f=}) -- (\ref{Tp-R}) to infer that $v_1(x,y)$ and $v_2(x,y)$ have identical scattering amplitudes;
    \be
    f^{l/r}_1(\theta)=f^{l/r}_2(\theta)~~~{\rm for}~~~k\in\sR.
    \label{condi-0}
    \ee
A quick examination of (\ref{H=}) shows that $\bH(x)$ has a linear dependence on $v$. This allows us to identify the condition $\bH_1(x)=\bH_2(x)$ with $\delta\bH(x)=\bzero$,
where $\delta\bH(x)$ is the Hamiltonian operator (\ref{H=}) with $\delta v:=v_2-v_1$ playing the role of $v$. This observation together with  Eqs.~(\ref{v-op}) and (\ref{H=}) prove the following result.
    \begin{itemize}
    \item[]{\em Theorem~2:} Let $\sR$ be a range of values of the incident wavenumber $k$. Then a pair of scattering potentials, $v_1(x,y)$ and $v_2(x,y)$, have identical scattering properties for $k\in\sR$, if the potential given by their difference, namely $\delta v(x,y):=v_2(x,y)-v_1(x,y)$, is invisible for $k\in\sR$.
    \end{itemize}

    If $v_2(x,y)$ is obtained from $v_1(x,y)$ by adding an extra piece, namely $\delta v(x,y)$, such that along the $x$-axis the support of $\delta v(x,y)$ lies to the left (respectively right) of that of $v_1(x,y)$, i.e., there is some $a\in\R$ such that  $\delta v(x,y)=0$ for $x>a$ and $v_1(x,y)=0$ for $x<a$ (respectively $\delta v(x,y)=0$ for $x<a$ and $v_1(x,y)=0$ for $x>a$), then the statement of Theorem~2 is rather trivial. It is also very easy to verify this statement for weak potentials where the first Born approximation is reliable \cite{prsa-2016}. Note however that Theorem~2 is a non-perturbative result holding for both weak and strong potentials with arbitrary supports. In particular, it applies to cases where $v_1(x,y)$ and $v_2(x,y)$ are strong potentials with supports of $\delta v(x,y)$ and $v_1(x,y)$ overlaping in one or several regions of space.  For these potentials the statement of Theorem~2 is highly nontrivial. The generality of this theorem and the simplicity of its proof signify the effectiveness of the transfer matrix formulation of the scattering theory in two dimensions.

Combining Theorems~1 and 2, we arrive at the following criterion for $\alpha$-equivalence.
    \begin{itemize}
    \item[]{\em Theorem~3:} Let $\alpha$ be a wavenumber scale. Then a pair of scattering potentials, $v_1(x,y)$ and $v_2(x,y)$, are $\alpha$-equivalent, if
    \be
    \tilde v_1(x,\fK_y)=\tilde v_2(x,\fK_y)~~~{\rm for}~~~\fK_y\leq2\alpha.
    \label{condi-1}
    \ee
    \end{itemize}
Condition~(\ref{condi-1}), which we can also state as: ``$\widetilde{\delta{v}}(x,\fK_y)=0$ for $\fK_y\leq2\alpha$,'' is equivalent to \cite{ol-2017}:
    \be
    v_2(x,y)=v_1(x,y)+ e^{2i\alpha y}u(x,y),
    \label{condi-2a}
    \ee
where $u(x,y)$ is a function whose Fourier transform with respect to $y$ vanishes in the negative $\fK_y$-axis, i.e.,
    \be
    \tilde u(x,\fK_y)=0~~~{\rm for}~~~\fK_y\leq 0.
    \label{condi-2b}
    \ee
For a fixed $x$, $u(x,y)$ belongs to the unidirectionally invisible potentials in one dimension that are studied in Refs.~\cite{horsley,jiang}. See also Refs.~\cite{longhi1,longhi2,longhi3,hl-review}.

It is easy to see that (\ref{condi-2b}) is equivalent to: $u(x,y)=\int_0^\infty d\fK_y\, e^{iy\fK_y}f(\fK_y)$, where $f:[0,\infty)\to\C$ is any function satisfying $\int_0^\infty d\fK_y\,|f(\fK_y)|<\infty$. We can use this observation together with (\ref{condi-2a}) to establish the following characterization of $\alpha$-equivalent potentials in two dimensions.
    \begin{itemize}
    \item[]{\em Theorem~4:} Let $\alpha$ be a wavenumber scale. Then a pair of scattering potentials, $v_1(x,y)$ and $v_2(x,y)$, are $\alpha$-equivalent, if there is a function $f:[0,\infty)\to\C$ such that $\int_0^\infty dq\,|f(q)|$ exists (is finite) and
    \be
    v_2(x,y)=v_1(x,y)+ e^{2i\alpha y}\int_0^\infty dq\, e^{iy q}f(q).
    \label{condi-01c}
    \ee
    \end{itemize}

Theorems 1-4 admit the following three-dimensional generalizations.
    \begin{itemize}
    \item[]{\em Theorem~5:} Let $v(x,y,z)$ be a scattering potential, and $\tilde v(\fK_x,\fK_y,z)$ be the Fourier transform of $v(x,y,z)$ with respect to $x$ and $y$. Then $v(x,y,z)$ is invisible for incident waves with wavenumber $k\leq\alpha$, if
    \be
    \tilde v(\fK_x,\fK_y,z)=0~~~{\rm for}~~~\sqrt{\fK_x^2+\fK_y^2}\leq 2\alpha.
    \label{condi-10}
    \ee

    \item[]{\em Theorem~6:} Let $\sR$ be a range of values of the incident wavenumber $k$. Then a pair of scattering potentials, $v_1(x,y,z)$ and $v_2(x,y,z)$, have identical scattering properties for $k\in\sR$, if the potential given by their difference, namely $\delta v(x,y,z):=v_1(x,y,z)-v_2(x,y,z)$, is invisible for $k\in\sR$.

    \item[]{\em Theorem~7:} A pair of scattering potentials, $v_1(x,y,z)$ and $v_2(x,y,z)$, are $\alpha$-equivalent, if
    \be
    \tilde v_1(\fK_x,\fK_y,z)=\tilde v_2(\fK_x,\fK_y,z)~~~{\rm for}~~~
    \fK_x\leq2\alpha~{\rm and}~\fK_y\leq2\alpha.
    \label{condi-11}
    \ee

    \item[]{\em Theorem~8:} A pair of scattering potentials, $v_1(x,y,z)$ and $v_2(x,y,z)$, are $\alpha$-equivalent, if there is function $f:[0,\infty)\times[0,\infty)\to\C$ such that $\int_0^\infty dq_x\int_0^\infty dq_y \,| f(q_x,q_y,z)|$ exists and
    \be
    v_2(x,y,z)=v_1(x,y,z)+
    e^{2i\alpha(x+y)}\int_0^\infty dq_x\int_0^\infty dq_y \;e^{i(xq_x+yq_y)} f(q_x,q_y,z).
    \label{v=v+u}
    \ee

    \end{itemize}

We can prove these theorems by pursuing the same approach that led us to the proof of Theorems 1-4. In particular, we make use of the following observations \cite{pra-2016,prsa-2016}:
    \begin{itemize}
    \item[] (i) The scattering amplitude for a scattering potential $v(x,y,z)$ may be expressed in terms of the entries of the corresponding transfer matrix. This is a $2\times 2$ matrix $\bM$ with operator entries acting in $\sF^1_k$, where $\sF^d_k$ is the space of test functions $\phi:\R^2\to\C^d$ vanishing outside the disk
$\sD_k:=\left\{\:\vec p\in\R^2\:\big|\: |\vec p|<k\right\}$, i.e.,
    \[\sF^d_k:=\left\{\:\phi:\R^2\to\C^d\:\big|\:\phi(\vec p)=\bzero~{\rm for}~
    |\vec p|\geq k^2\:\right\},\]
where $\vec p:=p_x\hat\bbe_x+p_y\hat\bbe_y$ and $\bzero$ stands for the zero element of $\C^d$.
    \item[](ii) $\bM$ may be expressed as the time-ordered exponential of an effective $k$-dependent Hamiltonian operator $\bH(z)$ acting in $\sF^2_k$, with $z$ playing the role of time, i.e., $\bM=\sT\exp\left[-i\int_{-\infty}^\infty dz\,\bH(z)\right]$, where for all $\Phi\in\sF^2_k$,
    \be
    [{\bH}(z)\Phi](\vec p):=\frac{1}{2\varpi(\vec p)}
    e^{-i\varpi(\vec p)z\boldsymbol{\sigma}_3}
    v(i\vec\partial_p,z)\,\boldsymbol{\cK}\,
    e^{i\varpi(\vec p)z\boldsymbol{\sigma}_3}\Phi(\vec p),
    \label{sH=3D}
    \ee
$v(i\vec\partial_p,z):=v(i\partial_{p_x},i\partial_{p_y},z)$ is the operator acting in $\sF^1_k$ according to
    \be
    \big[v(i\vec\partial_p,z)\phi\big](\vec p):=
    \frac{1}{4\pi^2}\int_{\sD_k}d^2\vec q\:\tilde v(\vec p-\vec q,z)\phi(\vec q),
    \label{tv-3d}
    \ee
and $\tilde v(\vec\fK,z):=\tilde v(\fK_x,\fK_y,z)$ is the Fourier transform of $v(x,y,z)$ with respect to $x$ and $y$.
    \item[](iii) $v(x,y,z)$ is invisible for a range of values of the wavenumber $k$, if $\bM$ coincides with the identity operator $\bI$ acting in $\sF^2_k$ for these values of $k$. The latter holds if $\bH(z)$ vanishes.
    \item[](iv) $\bH(z)$ is a linear function of the potential $v(x,y,z)$. In particular, if
    $\bH_1(z)$, $\bH_2(z)$, and $\delta\bH(z)$ are the Hamiltonians corresponding to potentials $v_1(x,y,z)$, $v_2(x,y,z)$, and $v_2(x,y,z)-v_1(x,y,z)$, then $\delta\bH(z)=\bH_2(z)-\bH_1(z)$.
    \item[](v) Condition (\ref{condi-11}) is equivalent to $v(x,y,z)=e^{2i\alpha(x+y)}u(x,y,z)$, where $u:\R^3\to\C$ is a function such that $\tilde u(\fK_x,\fK_y,z)=0$ for $\fK_x\leq 0$ and $\fK_y\leq 0$. This in turn implies that $u(x,y,z)=\int_0^\infty d\fK_x\int_0^\infty d\fK_y \;e^{i(x\fK_x+y\fK_y)} f(\fK_x,\fK_y,z)$ for some function $f:[0,\infty)\times[0,\infty)\to\C$ with finite $\int_0^\infty dq_x\int_0^\infty dq_y|f(q_x,q_y,z)|$.
    \end{itemize}
Theorem~5 is a consequence of (iii) and the fact that whenever (\ref{condi-10}) holds, the right-hand side of (\ref{tv-3d}) and consquently $\bH(z)$ vanishes for $k\leq\alpha$. Theorem~6 follows from (ii), (iii), and (iv). To prove Theorem~7, we note that $\fK_x\leq2\alpha$ and $\fK_y\leq2\alpha$ imply $\sqrt{\fK_x^2+\fK_y^2}\leq 2\alpha$ and make use of Theorem~5 to show that $\delta v:=v_2-v_1$ is invisible for $k\leq\alpha$. This together with Theorem~6 imply Theorem~7. Theorem~8 follows from (v) and Theorem~7.

As an example of the application of Theorem~8, consider taking
    \[f(q_x,q_y,z)=\tilde\fz\,q_x^{n_x}q_y^{n_y}e^{-(a_xq_x+a_yq_y)}e^{-z^2/2a_z^2},\]
where $\tilde\fz$ is a real or complex coupling constant, $a_x,a_y$, and $a_z$ are positive real parameters, and $n_x$ and $n_y$ are positive integers. Then, condition (\ref{v=v+u}) for the $\alpha$-equivalence of $v_1$ and $v_2$ takes the form:
    \be
    v_2(x,y,z)=v_1(x,y,z)+\frac{\fz\,e^{2i\alpha(x+y)}e^{-z^2/2a_z^2}}{
    \left(x/a_x+i\right)^{n_x+1}
    \left(y/a_y+i\right)^{n_y+1}},
    \label{w=2}
    \ee
where $\fz:=n_x!n_y!\,\tilde\fz/[(-ia_x)^{n_x+1}(-ia_y)^{n_y+1}]$.

In summary, we have investigated the application of the transfer-matrix formulation of scattering theory in the study of complex scattering potentials with identical scattering features for an extended range $\sR$ of values of the incident wave number. In particular, we have offered a rather general solution of this problem for the scattering of scalar waves in two and three dimensions whenever $\sR$ is a finite interval of the form $(0,\alpha]$. This is the problem of characterizing $\alpha$-equivalent complex potentials in two and three dimensions. Our solution is surprisingly simple and powerful in the sense that we can use it to construct large classes of $\alpha$-equivalent potentials without restricting their support or invoking perturbation theory. We attribute the simplicity and generality of this solution to the effectiveness of the transfer-matrix formulation of potential scattering in two and three dimensions \cite{pra-2016}. \vspace{6pt}

\noindent{\bf Acknowledgements:}
We thank Alexander Moroz for bringing Ref.~\cite{abeles} to our attention and 
Turkish Academy of Sciences (T\"UBA) for supporting FL's visit to Ko\c{c} University in 2018 during which this work was initiated. AM has been supported by T\"UBA's membership grant.

\ed